\documentclass[aps,prd,preprintnumbers,superscriptaddress,nofootinbib,showpacs,twocolumn]{revtex4}%
\usepackage[dvipdfmx]{graphicx}
\usepackage{bm,latexsym,amsmath,amssymb,amsfonts,mathrsfs}
\usepackage{color}
\input{colordvi.tex}
\usepackage[dvipdfmx]{hyperref}
\usepackage{url}
\hypersetup{
    colorlinks=true,
    citecolor=cyan,
}
\newcommand{\Equref}[1]{(\ref{#1})}
\newcommand{\henbi}{\partial}
\def\coloneqq{\mathrel{\mathop:}=}%
\begin{document}

\title{Instability of hairy black holes in shift-symmetric Horndeski theories}

\author{Hiromu~Ogawa}
\email[Email: ]{jh.ogawa"at"rikkyo.ac.jp}
\affiliation{Department of Physics, Rikkyo University, Toshima, Tokyo 171-8501, Japan
}
\author{Tsutomu~Kobayashi}
\email[Email: ]{tsutomu"at"rikkyo.ac.jp}
\affiliation{Department of Physics, Rikkyo University, Toshima, Tokyo 171-8501, Japan
}
\author{Teruaki~Suyama}
\email[Email: ]{suyama"at"resceu.s.u-tokyo.ac.jp}
\affiliation{Research Center for the Early Universe (RESCEU),
Graduate School of Science, The University of Tokyo, Tokyo 113-0033, Japan
}
\begin{abstract}
Recently it was pointed out that in shift-symmetric scalar-tensor theories
a black hole can have nontrivial scalar hair which depends linearly on time.
We develop black hole perturbation theory for such solutions and
compute the quadratic action of odd-parity perturbations.
We show that
around all the solutions known so far with such time-dependent scalar hair
the perturbations trigger instabilities or are presumably strongly coupled.
\end{abstract}
\preprint{RUP-15-22, RESCEU-27/15}
\maketitle

\section{Introduction}

Black holes are intriguing objects for many reasons.
From a theoretical viewpoint,
black holes have a rich mathematical structure
with regard to their uniqueness~\cite{Israel:1967wq,Israel:1967za} in and beyond general relativity in four spacetime dimensions.
The stability of black holes is also of great interest.
Black holes carry entropy and comply with the laws of thermodynamics~\cite{Bekenstein1,Bardeen:1973gs},
which is expected to shed light on quantum aspects of gravity.
From an astrophysical viewpoint, it is strongly suggested that there is
a black hole at the center of our Galaxy~\cite{Kormendy:1995er}.
Not only such supermassive black holes but also smaller
astrophysical ones are considered to be
promising sources of gravitational waves, the future detection of which would
provide us with unique information concerning astrophysics and gravitational physics.
Among those various aspects, in this paper,
we focus on the question as to whether black holes
can have nontrivial hair and whether they are stable in scalar-tensor theories of gravity.

The discovery of the accelerated expansion of the Universe~\cite{expan1,Riess:1998cb}
gives rise to a growing motivation for studying gravitational theories with a scalar field,
as it could be caused by a dynamical scalar field called dark energy~\cite{Copeland:2006wr}.
Alternatively, it is also argued that cosmic acceleration may be caused by
a modification of gravity~\cite{clifton}.
Modified gravity can be described at least effectively by
adding a scalar degree of freedom to the gravitational action.
Thus, theories composed of a metric and a scalar field are ubiquitous,
and it is important to explore aspects of scalar-tensor theories.

In many scalar-tensor theories, it has been shown that
black holes cannot support nontrivial scalar hair.
This was first pointed out in the context of the Brans-Dicke theory~\cite{Hawking:1972qk},
and later the theorem was extended to more general scalar-tensor theories~\cite{Sotiriou:2011dz}
including $k-$essence~\cite{Graham:2014mda,Graham:2014ina}.
It was also proven that a Galileon cannot develop a nontrivial configuration
around a static and spherically symmetric black hole~\cite{Hui}.
The proof relies essentially on the symmetry under a constant shift of the scalar field, $\phi\to\phi+c$,
and hence the same conclusion would hold true in general shift-symmetric scalar-tensor theories. 
See \cite{Herdeiro:2015waa} for a review.

The underlying assumption of
the no-hair proof of~\cite{Hui} is that the scalar field is static.
However, it was noticed that the scalar field can be linearly dependent on time
while the metric remains static in the shift-symmetric theories~\cite{Babichev}.\footnote{By
relaxing the asymptotic conditions one can construct hairy solutions
with a static scalar field~\cite{Rinaldi,Anabalon:2013oea,Minamitsuji1}. See~\cite{Sotiriou1,Sotiriou2} for yet another loophole.}
Utilizing this fact, the authors of~\cite{Babichev}
have worked out several exact black hole solutions dressed with nontrivial scalar hair
in the theory with the derivative coupling to the Einstein tensor.
Later, those solutions were generalized~\cite{Kobayashi:exa}
in a certain shift-symmetric subclass of the Horndeski theory~\cite{Horndeski,Deffayet,Kobayashi:ge}.
See~\cite{Charmousis:2014zaa} for a bi-scalar extension of the hairy solutions
and~\cite{Babichev:2015rva} for a charged generalization.

The purpose of the present work is to study the stability
of the hairy black hole solutions of~\cite{Babichev,Kobayashi:exa}.
To do so we compute the action rather than the equations of motion
for metric perturbations. The action approach was taken also in~\cite{DeFelice, Motohashi1,Motohashi:2011pw} 
to show the appearance of ghosts in some modified gravity theories.
Black hole perturbation theory in the action approach
has been formulated in the context of the Horndeski theory~\cite{Kobayashi:per1, Kobayashi:per2},
giving the useful stability conditions in terms of the arbitrary functions in the Horndeski action (see also \cite{Cisterna:2015uya}).
The results of~\cite{Kobayashi:per1, Kobayashi:per2} cannot, however, be
applied to the black hole solutions with time-dependent scalar hair,
as the background scalar field is assumed to be static in~\cite{Kobayashi:per1, Kobayashi:per2}.
We therefore extend the formulation of~\cite{Kobayashi:per1, Kobayashi:per2}
to allow for the linear time dependence of the background scalar
within the shift-symmetric subclass of the Horndeski theory.

The organization of this paper is as follows.
In Sec. II, we review the black hole solutions with time-dependent scalar hair
in the Horndeski theory, following~\cite{Kobayashi:exa}.
In Sec. III, we compute the quadratic action for odd parity perturbations
and give the stability conditions.
In Sec. IV we apply the stability conditions to the hairy black holes
and draw our conclusions on their stability.

\section{Hairy black holes in shift-symmetric scalar-tensor theories}

\subsection{Shift (and reflection) symmetric scalar-tensor theories}

The Horndeski theory is the most general scalar-tensor theory with
second-order field equations in four dimensions~\cite{Horndeski}.
We work in the subclass of the Horndeski theory having symmetries under
$\phi\to\phi+c$ and $\phi\to-\phi$.
The action we consider is thus given by
\begin{equation}
S=\int d^4x\sqrt{-g}(\mathcal{L}_2+\mathcal{L}_4),\label{eq:exact}
\end{equation}
where
\begin{align}
\mathcal{L}_2&=G_2(X),\\
\mathcal{L}_4&=G_4(X)R+G_{4X}\left[ (\Box \phi)^2-(\nabla_{\mu}\nabla_{\nu}\phi)^2\right].
\end{align}
Here, $G_2$ and $G_4$ are arbitrary functions of $X\coloneqq -(\henbi \phi)^2/2$,
$R$ is the Ricci scalar, and we use the notation 
$G_{4X}\coloneqq \henbi G_{4}/\henbi X$.
This is the most general second-order scalar-tensor theory
having shift and reflection symmetries.
The shift symmetry is essential for black holes to have time-dependent scalar hair.
However, the reflection symmetry is rather thought of as a simplifying assumption
used to remove the so-called ${\cal L}_3$ and ${\cal L}_5$ terms from the theory.

The theory with a nonminimal derivative coupling to the Einstein tensor,
\begin{align}
S=\int d^4x\sqrt{-g}\left[\zeta R-\eta (\henbi \phi)^2+\beta G^{\mu\nu}\henbi_{\mu}\phi\henbi_{\nu}\phi -2\Lambda
\right],\label{nmdcgmn}
\end{align}
with $\zeta$, $\eta$, $\beta$ and $\Lambda$ being constants,
is often considered in the literature (see, {\it e.g.,}~\cite{Rinaldi,Gubitosi:2011sg,Germani,Minamitsuji1}).
This theory is a specific case of the general theory \Equref{eq:exact},
as is confirmed by taking
\begin{align}
G_2=-2\Lambda+2\eta X,\quad G_4=\zeta+\beta X, 
\end{align}
and performing integration by parts.

\subsection{Black holes with time-dependent scalar hair}

We consider a static and spherically symmetric metric as a background solution.
The background metric $\bar{g}_{\mu\nu}$ is therefore of the form
\begin{align}
\bar{g}_{\mu\nu}dx^{\mu}dx^{\nu}
 =-A(r)dt^2+\frac{dr^2}{B(r)}+r^2d\Omega^2,\label{eq:background}
\end{align}
where $d\Omega^2=d\theta^2+\sin^2\theta d\varphi^2$.
The scalar field can, however,  be dependent linearly on time
as a consequence of the shift symmetry,
\begin{equation}
\phi(r,t)=qt+\psi(r),\label{sc-ansatz}
\end{equation}	
where $q$ is a constant, and nonvanishing $q$ is crucial for
a hairy solution~\cite{Babichev}.

Several exact hairy black hole solutions have been obtained
in the theories with the specific action~\Equref{nmdcgmn} and the general one~\Equref{eq:exact}.
To study the stability of those solutions, we formulate the black hole perturbation theory
with time-dependent scalar hair in the theory~\Equref{eq:exact}.
To do so, we do not rely on the concrete form of the solution known so far
but only use the assumption that the metric and the scalar field are of the form~\Equref{eq:background}
and~\Equref{sc-ansatz} and satisfy the background field equations.

The background field equations can be derived as follows.
We substitute the metric
\begin{align}
\bar{g}_{\mu\nu}dx^{\mu}dx^{\nu}
=&-A(r)dt^2+\frac{dr^2}{B(r)}+2C(r)dtdr
\notag \\&
+D(r)r^2d\Omega^2,\label{eq:background1}
\end{align}
and the scalar field~\Equref{sc-ansatz} to the action~\Equref{eq:exact}.
Varying the action with respect to $A$, $B$, $C$, and $D$,
and then setting $C=0$ and $D=1$,
we obtain the gravitational field equations.
We write the corresponding equations as
\begin{align}
\mathcal{E}_{A}&=0,\quad \mathcal{E}_{B}=0,\quad \mathcal{E}_{C}=0,\quad \mathcal{E}_{D}=0,
\label{eq:backgroundeq}
\end{align}
the explicit forms of which are found in Appendix A.
One can also derive the scalar-field equation of motion by varying the action
with respect to $\psi$. 
In formulating black hole perturbation theory below, use of those equations is
crucial to simplify the second order action.
For this purpose, it is useful to know that not all of the equations are independent.
Actually, thanks to the Bianchi identities,
$\mathcal{E}_D=0$ and the scalar-field equation are automatically satisfied
once the other equations are assumed.

Let us now summarize the hairy black hole solutions obtained so far in the literature.
The solutions can be classified into several groups.
The key equation for the classification is
\begin{eqnarray}
\frac{d}{dr}\left[X{\cal H}(X)\left(1-r^2F(X)\right)\right] = 0,\label{keytocl}
\end{eqnarray}
which can be derived by combining the gravitational field equations~\Equref{eq:backgroundeq}
under the assumption that $q\neq 0$~\cite{Kobayashi:exa}.
Here we defined, for convenience
\begin{align}
{\cal H}(X):=&\;2(G_4-2XG_{4X}),\label{1defH}
\\
F(X):=&\;\frac{\henbi_{X} (G_2\mathcal{H})}{8X(G_{4X}^2+G_4G_{4XX})}.
\end{align}
Equation~\Equref{keytocl} can be integrated to give
an algebraic equation for $X$:
\begin{align}
X{\cal H}\left(1-r^2F\right)={\rm const}.
\end{align}
From this equation, one can determine $X=X(r)$,
which, in general, depends on the concrete form of $G_2$ and $G_4$, and also on the
integration constant in the right hand side.
One can then use the other components of the gravitational field equations
to fix $A=A(r)$ and $B=B(r)$.
Note, however, that the procedure to find exact solutions goes
almost independently of the concrete form of $G_2$ and $G_4$
if one focuses on the special cases where $F(X)=0$ or ${\cal H}(X)=0$ is fulfilled.
Accordingly, $X=$ const for such solutions.
Actually, it turns out that all the known solutions satisfy either $F(X)=0$ or ${\cal H}(X)=0$
as displayed below.

\subsubsection{$F(X)=0$}\label{F0sol}

With an appropriate rescaling of the time coordinate,
we obtain the following solution for $F(X)=0$:
\begin{align}
A=B=1-\frac{\mu}{r}-\frac{\Lambda_{\rm eff}}{3}r^2,
\end{align}
where $\mu$ is an integration constant and
\begin{align}
\Lambda_{\rm eff}(X):= -\frac{1}{2}\frac{G_2G_{4XX}+G_{2X}G_{4X}}{G_{4X}^2+G_{4}G_{4XX}}
\;\left(={\rm const}\right).
\end{align}
Interestingly, the metric is identical to the Schwarzschild-de Sitter solution in general relativity,
though in this case the scalar field exhibits a nontrivial profile. Note that
the effective cosmological constant has nothing to do with
the true cosmological constant.
This solution was first discovered in the theory~\Equref{nmdcgmn} in~\cite{Babichev}
and then was generalized in the context of the Horndeski theory~\Equref{eq:exact}
in~\cite{Kobayashi:exa}.

\subsubsection{${\cal H}(X)=0$}\label{H0sol}

The solution satisfying ${\cal H}(X)=0$ is more complicated and is given by
\begin{align}
A=&\;\frac{\Lambda_{\rm eff}}{F}+\left(1-\frac{\Lambda_{\rm eff}}{F}\right){\cal T}(r)-\frac{\mu}{r},
\\
B=&\;(1-Fr^2)A,
\end{align}
with
\begin{equation}
\mathcal{T}(r)\coloneqq 
\begin{cases}
\dfrac{1}{2\sqrt{F}r}\ln{\left| \dfrac{1+\sqrt{F}r }{1-\sqrt{F}r }\right|}&(F>0)\\
\dfrac{\arctan{(\sqrt{-F}r)}}{\sqrt{-F}r}&(F <0)
\end{cases},
\end{equation}
where $\mu$ is an integration constant and
an appropriate rescaling of the time coordinate is understood.
The above solution includes the Schwarzschild black hole in an Einstein static universe~\cite{Babichev}
as a special case.

\section{Perturbations of a black hole with time-dependent scalar hair}

We now calculate the quadratic action for metric perturbations.
Regge and Wheeler~\cite{Regge} and Zerilli~\cite{Zerilli} were
the first to develop black hole perturbation theory.
We follow their works and decompose the metric perturbations
into odd and even modes according to their transformation properties under
two-dimensional rotation. In this paper, we only consider the odd parity sector
in which the scalar field does not acquire fluctuations.
While Regge, Wheeler, and Zerilli worked out the equations of motion
for the master variables, we compute the quadratic action for the perturbations~\cite{Moncrief:1974am}.
By taking the action approach, one can derive stability conditions
in a transparent manner. Looking at the quadratic action is particularly important in
modified gravity theories, because one must care about the sign of the kinetic terms
of the dynamical variables and associated ghost instabilities.
In~\cite{Kobayashi:per1,Kobayashi:per2},
the quadratic actions for odd and even parity perturbations around
a static and spherically symmetric background
are derived from the Horndeski theory, making the underlying assumption that
the background scalar field is also static.
We now extend the work of~\cite{Kobayashi:per1,Kobayashi:per2}
to admit time-dependent scalar hair, following closely the previous works of~\cite{DeFelice,Motohashi1}
and~\cite{Kobayashi:per1,Kobayashi:per2}.

\subsection{Odd parity metric perturbations}

The metric perturbations $h_{\mu\nu}=g_{\mu\nu}-\bar{g}_{\mu\nu}$
in the odd parity sector can be written using spherical harmonics $Y_{lm}(\theta,\varphi)$
as
\begin{align}
h_{tt}&=0,\quad h_{tr}=0,\quad h_{rr}=0,\notag \\
h_{ta}&=\sum_{l,m}^{}h_{0,lm}(t,r)E_{ab}\henbi^{b}Y_{lm}(\theta,\varphi),\notag\\
h_{ra}&=\sum_{l,m}^{}h_{1,lm}(t,r)E_{ab}\henbi^{b}Y_{lm}(\theta,\varphi),\notag\\
h_{ab}&=\frac{1}{2}\sum_{l,m}^{}h_{2,lm}(t,r)[{E_a}^c\nabla_c\nabla_bY_{lm}(\theta,\varphi)\notag\\
&\hspace{2cm}+{E_b}^c\nabla_c\nabla_aY_{lm}(\theta,\varphi)],
\end{align}
where $E_{ab}\coloneqq\sqrt{\det{\gamma}}\varepsilon_{ab}$ with $\gamma_{ab}$ being the 
two-dimensional metric on a unit sphere and $\varepsilon_{ab}$ being  
the totally anti-symmetric symbol with $\varepsilon_{\theta\varphi}=1$.

Not all the above perturbation variables are physical;
by performing the gauge transformation $x^a\to x^a+\xi^a$, 
where
\begin{equation}
\xi_{a}=\sum_{l,m}^{}\Lambda_{lm}(t,r){E_{a}}^{b}\nabla_bY_{lm},
\end{equation}
with an appropriate choice of $\Lambda_{lm}$,
one can eliminate $h_2$, leaving $h_0$ and $h_1$.
This is the Regge-Wheeler gauge, which is always possible for $l\ge 2$ modes. However,
for the dipole perturbations ($l=1$),
$h_{ab}$ vanishes identically and the gauge is not fixed completely.
We therefore consider the dipole mode separately.
Note that there are no odd parity perturbations for $l=0$.

\subsection{Quadratic action for odd parity perturbations with $l\ge2$}

Substituting the perturbed metric
into the action~\Equref{eq:exact} and expanding it to second order, we obtain 
the quadratic action for the odd parity perturbations. After
lengthy but straightforward calculations we arrive at, for a given set of $(l,m)$,\footnote{Since
different $(l,m)$ modes do not mix, one can treat each single $(l,m)$ separately.
Thanks to the spherical symmetry, the action is independent of $m$, and hence,
one may set $m=0$ from the beginning without loss of generality.
We omit the subscripts $l,m$ when unnecessary.}
\begin{align}
S^{(2)}=\int_{}^{}dtdr\mathcal{L}^{(2)}, \label{2nd-action}
\end{align}
with
\begin{align}
\frac{2l+1}{2\pi}\mathcal{L}^{(2)}=&\; a_1h_0^2+a_2h_1^2+a_3\biggl(\dot{h}_1^2-2
   h_0' \dot{h}_1\notag\\
   & +h_0'^2+\frac{4
   h_0
   \dot{h}_1}{r}\biggr)+a_4
   h_0h_1,\label{eq:quadlag}
\end{align}
where a dot (a prime) stands for differentiation with respect to $t$ ($r$).
The coefficients $a_1$, $a_2$, $a_3$, and $a_4$ 
are given by
\begin{align}
a_1&=\frac{l(l+1)}{r^2}\left[ \frac{d}{dr}\left( r\sqrt{\frac{B}{A}}\mathcal{H}\right)
+\frac{(l-1)(l+2)}{2\sqrt{AB}}\mathcal{F}\right],\label{eq:a1}\\
a_2&=-\frac{(l-1)l(l+1)(l+2)}{2}\frac{\sqrt{AB}}{r^2}\mathcal{G},\label{eq:a2}\\
a_3&=\frac{l(l+1)}{2}\sqrt{\frac{B}{A}}\mathcal{H},\label{eq:a3}\\
a_4&= \frac{(l-1)l(l+1)(l+2)}{r^2}\sqrt{\frac{B}{A}}\mathcal{J}\label{eq:a4},
\end{align}
where we defined
\begin{align}
\mathcal{F}&= 2 \left( G_4-\frac{q^2}{A}G_{4X}\right),\label{eq:FF}\\
\mathcal{G}&= 2 \left( G_4-2XG_{4X}+\frac{q^2}{A}G_{4X} \right),\label{eq:GG}\\
\mathcal{H}&=2 \left( G_4-2XG_{4X}\right),\label{eq:HH}\\
\mathcal{J}&=2qG_{4X}\psi',\label{eq:JJ}
\end{align}
and used the background equations \Equref{eq:backgroundeq} to simplify the expressions.
Note here that ${\cal H}$ is the same as the one introduced earlier in Eq.~\Equref{1defH}.
We see that $q$ appears in several places in the coefficients;
in particular, the term $a_4h_0h_1$ arises due to nonvanishing $q$ and hence is completely new.

The Lagrangian shows that $h_1$ is dynamical while $h_0$ is not.
Thus, variation with respect to $h_0$ yields a constraint equation.
However, we cannot solve straightforwardly the constraint for $h_0$ because
it contains $h_0'$.
To remove the nondynamical degree of 
freedom from the Lagrangian, we instead rewrite the Lagrangian
by introducing an auxiliary field $\chi$
as~\cite{DeFelice,Motohashi1}
\begin{align}
\frac{2l+1}{2\pi}\mathcal{L}^{(2)}&=\left[a_1-\frac{2(ra_3)'}{r^2}\right]h_0^2+a_2h_1^2\notag\\
&+a_3\left[-\chi^2+2\chi \left(\dot{h}_1-
   h_0' +\frac{2}{r}h_0\right)\right]+a_4
   h_0h_1.\label{eq:newLag}
\end{align}
It is easy to verify that Eq.~\Equref{eq:quadlag} is recovered
by eliminating the auxiliary field $\chi$ from \Equref{eq:newLag}.
Varying the new Lagrangian~\Equref{eq:newLag} with respect to $h_0$ and $h_1$,
one obtains the two equations that can now be solved for $h_0$ and $h_1$, giving
\begin{align}
h_0&=-\frac{2 r \left\{2 {a_2} \left[r (\chi{a_3})' +2\chi a_3\right]+r   \dot{\chi}{a_3} {a_4}
 \right\}}{4 a_2[r^2 {a_1}-2\left(ra_3\right)']-r^2 {a_4}^2},\label{h0=}\\
h_1&=\frac{4 {a_3} \dot{\chi} [r^2 {a_1}-2(ra_3)']+2 r a_4[ r(\chi{a_3})' +2a_3\chi]}{4 a_2[r^2 {a_1}-2(ra_3)']-r^2 {a_4}^2}.\label{eq:h0h1rel}
\end{align}
Substituting Eqs.~(\ref{h0=}) and (\ref{eq:h0h1rel}) back into Eq.~\Equref{eq:newLag},
we finally find the quadratic Lagrangian written in terms of only the dynamical variable $\chi$,
\begin{align}
\frac{2l+1}{2\pi}\mathcal{L}^{(2)}=&\;
\frac{l(l+1)}{2(l-1)(l+2)}\sqrt{\frac{B}{A}}\bigl[{b_1} \dot{\chi}^2-{b_2} \chi'^2\notag\\
&+{b_3} \dot{\chi}\chi' -l(l+1){b_4}\chi^2-V\chi^2\bigr],
\label{eq:newLag2}
\end{align}
where the coefficients $b_i$ are given by
\begin{align}
b_1&=\frac{r^2\mathcal{F}\mathcal{H}^2}{A\mathcal{F}\mathcal{G}+B\mathcal{J}^2}, \\
b_2&=\frac{r^2AB\mathcal{G}\mathcal{H}^2}{A\mathcal{F}\mathcal{G}+B\mathcal{J}^2},\\
b_3&=\frac{2r^2B\mathcal{H}^2\mathcal{J}}{A\mathcal{F}\mathcal{G}+B\mathcal{J}^2},\\
b_4&=\mathcal{H}.
\end{align}
The explicit form of the effective potential $V$ is given in Appendix B because of its long expression.
Once a solution to the Euler-Lagrange equation for $\chi$ is given,
one can use Eqs.~(\ref{h0=}) and~(\ref{eq:h0h1rel}) to determine $h_0$ and $h_1$,
and thus can fix all the perturbation variables in the Regge-Wheeler gauge.

A comment is now in order.
If ${\cal H}=0$, all the coefficients in the Lagrangian~(\ref{eq:newLag2}) vanish.
This implies
that fluctuations are strongly coupled around the background solutions with ${\cal H}=0$.

\subsection{Stability conditions}

To simplify the equations, let us suppress the $l$-dependent factor in Eq.~\Equref{eq:newLag2},
which is unimportant for the stability,
and consider the
Lagrangian density,
\begin{align}
\widetilde{\cal L}=\frac{1}{2}
\sqrt{\frac{B}{A}}\left\{{b_1} \dot{\chi}^2-{b_2} \chi'^2
+{b_3} \dot{\chi}\chi' -\left[l(l+1)b_4+V\right]\chi^2\right\}.
\end{align}
One can define the conjugate momentum as $\pi=\partial\widetilde{\cal L}/\partial \dot\chi$,
and then the Hamiltonian is given by
\begin{align}
H=\frac{1}{2}\int dr\sqrt{\frac{B}{A}}&\Biggl\{
\frac{1}{b_1}\left(\sqrt{\frac{A}{B}}\pi -\frac{1}{2} b_3\chi'\right)^2
\notag \\ & 
+b_2\chi'^2+\left[l(l+1)b_4+V\right]\chi^2
\Biggr\}.
\end{align}
In order for the first and second terms to be positive, it is required that
\begin{eqnarray}
b_1>0,\quad b_2>0.
\end{eqnarray}
Those ensure the positivity of the kinetic and radial gradient energies, respectively.
To avoid instabilities of large $l$ modes,
\begin{eqnarray}
b_4>0
\end{eqnarray}
is required as well.
Those three conditions are equivalent to
\begin{eqnarray}
{\cal F}>0,\quad{\cal G}>0, \quad{\cal H}>0.\label{stfgh}
\end{eqnarray}
Equation~\Equref{stfgh} can be used for any
black hole solutions with linearly time-dependent scalar hair
in a sufficiently wide subclass of the Horndeski theory \Equref{eq:exact}
and thus generalizes the result of~\cite{Kobayashi:per1}.

\subsection{Dipole perturbation}

The procedure to derive the quadratic action~\Equref{eq:newLag2} for the master variable
is not well-defined for the $l =1$ modes since division by $(l-1)$ is 
involved in the intermediate computations.
This necessitates doing the perturbation analysis for the dipole modes separately,
which is the purpose of this subsection.
To this end, we start with the action \Equref{2nd-action}.
This is nothing but the action expanded to second
order in terms of the original perturbation variables and is valid even for the dipole modes.
Plugging $l=1$ into the background-dependent coefficients $a_i\;(i=1,\cdots, 4)$,
we find $a_2$ and $a_4$ become zero and end up with
\begin{equation}
\frac{3}{2\pi}\mathcal{L}^{(2)}= \frac{2}{r^2} {(ra_3)}' h_0^2+a_3\biggl(\dot{h}_1^2-2
   h_0' \dot{h}_1 +h_0'^2+\frac{4 h_0 \dot{h}_1}{r}\biggr),
\end{equation} 
where $a_3=\mathcal{H}\sqrt{B/A}$.
As is the case in the higher multipole modes, the quadratic Lagrangian
becomes zero when $\mathcal{H}=0$,
which signals the strong coupling of the system.
In the below, we assume $\mathcal{H} \neq 0$.
Contrary to the higher multipole modes, the angular components of the metric perturbation
automatically vanish for the dipole modes and the above Lagrangian is valid in any gauge~\cite{Motohashi:2011pw}.
Indeed, we can explicitly verify the invariance of the Lagrangian (up to the total derivative)
under the gauge transformation,
\begin{equation}
h_0 \to h_0+{\dot \Lambda},~~~~~h_1 \to h_1+{\Lambda}'-\frac{2}{r}\Lambda,
\end{equation}
where $\Lambda (t,r)$ is an arbitrary function specifying the gauge transformation
for the angular coordinates.
By using this gauge degree of freedom, we can always choose a gauge
for which $h_1=0$.
We impose this gauge condition after we obtain the Euler-Lagrange equations for
$h_0$ and $h_1$.
There is still a gauge degree of freedom for $\Lambda$ such that $\Lambda =C(t) r^2$,
where $C(t)$ is an arbitrary function of $t$.
This residual gauge is used later to eliminate the gauge mode appearing in $h_0$.

The Euler-Lagrange equations for $h_0$ and $h_1$ take the forms of
\begin{align}
&\dot{h}_0'-\frac{2}{r}\dot{h}_0=0,\\
&a_3h_0''+a_3'h_0'-\frac{2(ra_3)'}{r^2}h_0=0.
\end{align}
Clearly, these equations look exactly the same as those in the case of the time independent 
scalar field ($q=0$) derived in~\cite{Kobayashi:per1}.
Dependence of the second equation on $q$ is only through $X$ contained in $\mathcal{H}$.
As a result, the solution takes exactly the same form as the one given in~\cite{Kobayashi:per1},
which is given as
\begin{equation}
h_0=\frac{3Jr^2}{4\pi}\int_{}^{r}\frac{d\bar{r}}{\bar{r}^4\mathcal{H}}\sqrt{\frac{A}{B}}+{\tilde C}(t)r^2,
\label{sol-h0}
\end{equation}
where $J$ is an integration constant, and ${\tilde C}(t)$ is an arbitrary function of $t$. 
As mentioned before, the second term is a gauge mode and can be removed by
setting $\Lambda=-\int^t {\tilde C}(t')dt' r^2$. 
The first term cannot be eliminated by any gauge transformation.
Physically, this represents a metric of the slowly rotating black hole,
that is, metric to first order in its angular momentum $J$.  
For a class of solutions with $F(X)=0$ discussed in Sec.~\ref{F0sol},
both $X$ and $A/B$ are independent of $r$.
Then, we have
\begin{equation}
h_0=-\frac{J}{4\pi \mathcal{H}(X)r},
\end{equation}
where $X$ here is a solution of $F(X)=0$.
The metric $h_0$ takes exactly the same form as the Kerr metric expanded
up to first order in the angular momentum.
This result is consistent with a recent study~\cite{Maselli:2015yva} in which
it is shown by a different approach that
the metric around a slowly rotating black hole is
identical to
that in general relativity for a wide class of shift-symmetric Horndeski theories.

\section{Discussion and conclusions}

In the previous section, we have obtained the stability conditions
for a black hole dressed with a time-dependent scalar field.
The main conclusion derived from the stability conditions ${\cal F}>0$ and ${\cal G}>0$
is as follows: for the solutions with $X=$ const, one has
\begin{align}
{\cal F}{\cal G}\simeq -4\left(\frac{q^2}{A}G_{4X}\right)^2<0,
\end{align}
in the vicinity of the horizon ($A=0$), and thus, either ${\cal F}$ or ${\cal G}$
is negative there, indicating that {\em the black hole solutions with $X=$ const
are unstable.}
This result holds in the general shift-symmetric theory~\Equref{eq:exact} and
can be applied to the Schwarzschild-de Sitter black hole
with time-dependent scalar hair presented in Sec.~\ref{F0sol}.
Since the  stability conditions are independent of $G_2$,
the stealth Schwarzschild solution
found in $G_2=0$ theories~\cite{Babichev, Kobayashi:exa} is also unstable.
The same instability occurs in the vicinity of the cosmological horizon.
In particular, this is true even for the special case of the de Sitter solution
(with time-dependent hair).

The instability we found in this paper manifests in the short-wavelength perturbations
localized sufficiently close to (but not on) the horizon,
because the stability conditions are violated in a local domain outside the horizon.
This local nature of the instability implies that it is not a coordinate artifact
(see Appendix C).
Nor is it due to the apparent singular behavior of the metric and the scalar field at
the horizon as the instability is developed not on the horizon.

Let us give a further discussion to support this conclusion in the following.
We can express a physically sensible perturbation as
\begin{equation}
\chi(t,r_*)=\sum_\omega e^{i\omega t} \chi_\omega (r_*),\label{eq:wave}
\end{equation}
where $\chi_\omega (r_*)$ forms a complete set and $r_*$ is
the tortoise coordinate defined as $r_*=\int dr/\sqrt{AB}$.
We claim that the complete set includes the mode whose $\omega$ is 
purely imaginary with a sufficiently large absolute value.
To see this, suppose that one prepares a sufficiently small-size wave packet 
$\sum_\omega \chi_\omega (r_*)$ sufficiently near the horizon at the initial time $t=0$.
Since it is localized, the square integral can be made finite, namely $\int dr_* {|\chi (r_*)|}^2 < \infty$.
Since the size of the packet is sufficiently small, we may consider it as a superposition
of plane waves and the majority of them have sufficiently large wavenumbers.
From the Euler-Lagrange equation for $\chi$, we find that $\omega$ corresponding to such a large
wavenumber $k$ is given by $\pm i\sqrt{A/B} k$ to a good approximation.
Thus, a sufficiently localized wave packet near the horizon contains $\chi_\omega (r_*)$ with
a purely imaginary $\omega$ having a very large absolute value.
This wave packet evolves subject to Eq.~(\ref{eq:wave}) and undergoes
rapid exponential growth.
As a result, the square integral $\int dr_* |\chi (t, r_*)|^2$ also
exhibits exponential growth
within a very short time-scale.
This is the nature of the ghost/gradient instability arising
due to the sign flip in the coefficients in the quadratic Lagrangian.
Based on this observation, we conclude that the black holes considered in this paper 
are classically unstable.

So far, we have made a purely classical argument.
We would like to emphasize that there is another
consequence from the
derived quadratic Lagrangian
leading to a more serious problem.
Since quantum mechanics describes our nature, in order for a
black hole solution to exist stably, the solution must be a ground state at least
locally in the state space.
As we have found, the Hamiltonian density of the metric perturbations is not bounded
from below in the vicinity of the horizon.
Since gravity is coupled to all matter fields that have excitations only with positive energy,
the unbounded Hamiltonian inevitably leads to instantaneous and enormous creation of matter 
and gravitational excitations near the horizon.
This pathological process, yet abiding by the energy conservation, 
inevitably takes place since any allowed process happens in quantum mechanics.
Thus, the black hole cannot be stable quantum mechanically
even if we do not allow 
any types of classical perturbations including the ones we mention above.
In this sense, the latter pathological instability is more serious than any classical
instabilities.
This point is well known in the field of modified gravity,
and a detailed discussion is given, for instance, in Ref.~\cite{Woodard:2015zca}.

Finally, a caveat must be given here.
For the solutions with ${\cal H}=0$ given in Sec.~\ref{H0sol},
the quadratic action vanishes, which implies that
the odd parity perturbations are strongly coupled on this background.
This example must be investigated carefully, but in any case is beyond the scope of the present paper.

In summary, we have derived the quadratic action for odd parity perturbations
of a black hole dressed with time-dependent scalar hair
in a sufficiently wide subclass of the Horndeski theory.
Based on the quadratic action, we have shown that the
perturbations of all the hairy black hole solutions with $X=$ const known so far
are either unstable or presumably strongly coupled.

\acknowledgments 
This work was supported in part by the JSPS Grant-in-Aid for Young 
Scientists (B) No.~24740161 (T.K.) and~No.~15K17632 (T.S.), MEXT Grant-in-Aid for Scientific Research on
Innovative Areas New Developments in Astrophysics
Through Multi-Messenger Observations of Gravitational
Wave Sources No.~15H00777 (T.S.) and MEXT KAKENHI Grant No. 15H05888 (T.K. and T.S.).

\appendix

\section{Background equations}
We define the background equations~\Equref{eq:backgroundeq} following~\cite{Maselli:2015yva}.
The gravitational field equations
can be divided into the $q$-independent part and the terms proportional to $q^2$
as
\begin{equation}
\mathcal{E}_{a}\coloneqq \mathcal{E}_{a}^{(0)}+\frac{q^2}{A}\mathcal{E}_{a}^{(t)},
\end{equation}
where $a=A,\;B,\;C,\;D$.
\begin{widetext}%
Explicitly,
\begin{flalign}
\mathcal{E}_{A}^{(0)}&=G_2-\frac{2}{r}\left(B'+\frac{B-1}{r}\right)G_4-\frac{2 B^2 \psi '}{r} \left(\frac{\psi'}{r}+\frac{2B'}{B}\psi'+2\psi''\right)G_{4X}+\frac{2B^3\psi'^{3}}{r}\left( \frac{B'}{B}\psi'+2\psi''\right)G_{4XX},&\\
\mathcal{E}_{A}^{(t)}&=-G_{2X}+\frac{2}{r^2}\left(-1+B+rB'\right)G_{4X}-\frac{2B^2\psi'}{r}\left( \frac{\psi'}{r}+\frac{B'}{B}\psi'+2\psi''\right)G_{4XX},&
\\
\mathcal{E}_{B}^{(0)}&=G_{2}+B\psi'^2G_{2X}-\frac{2}{r}\left( \frac{A'}{A}B+\frac{B-1}{r}\right)
G_{4}-\frac{2B\psi'^2}{r}\left(\frac{2A'}{A}B+\frac{2B-1}{r} \right)G_{4X}+\frac{2B^3\psi'^4}{r}\left( \frac{1}{r}+\frac{A'}{A}\right)G_{4XX},
& \\
\mathcal{E}_{B}^{(t)}&=\frac{2B}{r}\frac{A'}{A}G_{4X}-\frac{2B^2\psi'^2}{r}\frac{A'}{A}G_{4XX},&
\\
\mathcal{E}_{C}^{(0)}&=G_{2X}-\frac{2}{r}\left( \frac{B-1}{r}+\frac{A'}{A}B\right)G_{4X}+\frac{2}{r}B^2\psi'^2\left(\frac{1}{r}+\frac{A'}{A}\right)G_{4XX},&\\
\mathcal{E}_{C}^{(t)}&=-\frac{2}{r}\frac{A'}{A}BG_{4XX},& 
\\
\mathcal{E}_{D}^{(0)}&=G_2-\left[ \frac{1}{r}\sqrt{\frac{B}{A}}\left(r\sqrt{\frac{B}{A}}A'\right)'+\frac{B'}{r}\right]G_{4}-\frac{1}{2}B^2\psi'^2\left( \frac{4}{r}\frac{B'}{B}+\frac{2(A'+rA'')}{rA}-\frac{A'^2}{A^2}+2\frac{A'}{A}\frac{B'}{B}+\frac{4}{r}\frac{\psi''}{\psi'}+2\frac{A'}{A}\frac{\psi''}{\psi'}\right)G_{4X}\notag\\
&+\frac{1}{2}B^3\psi'^4\left( \frac{2}{r}+\frac{A'}{A}\right)\left( \frac{B'}{B}+2\frac{\psi''}{\psi'}\right)G_{4XX},\\
\mathcal{E}_{D}^{(t)}&=\left[ \frac{1}{r}\sqrt{\frac{B}{A}}\left(r\sqrt{\frac{B}{A}}A'\right)'-\frac{B}{2}\frac{A'^2}{A^2}\right]G_{4X}+\left[\frac{1}{2}\frac{A'}{A}B^2\psi'^2\left( \frac{2}{r}+\frac{A'}{A}-\frac{B'}{B}-2\frac{\psi''}{\psi}\right)-\frac{1}{2}\frac{A'^2}{A^3}Bq^2\right]G_{4XX}.
\end{flalign}

\section{Effective potential}\label{eff}

The effective potential in Eq.~\Equref{eq:newLag2} is given by
\begin{align}
V(r)=&\;\frac{\mathcal{H}}{ 2\left(A \mathcal{F} \mathcal{G}+B \mathcal{J}^2\right)^2}(-B(r^2 \mathcal{G} \mathcal{H} A' (\mathcal{F} \mathcal{G} A'+\mathcal{J}^2 B')\notag\\
&+B \mathcal{J} (2 r^2
   \mathcal{G} \mathcal{H} A' \mathcal{J}'+r \mathcal{J} (\mathcal{G} (-r \mathcal{H} A''+r A' \mathcal{H}'+4 \mathcal{H} A')-r \mathcal{H} A'
   \mathcal{G}')+4 \mathcal{J}^3))\notag\\
&-A^2\mathcal{G}^{2}(-r \mathcal{F}' (r \mathcal{H} B'+2 B (r \mathcal{H}'+2 \mathcal{H}))+\mathcal{F}
   (r (r \mathcal{H} B''+3 r B' \mathcal{H}'+4 \mathcal{H} B')+B (2 r^2 \mathcal{H}''+4 r \mathcal{H}'-4 \mathcal{H}))+4
   \mathcal{F}^2)\notag\\
&+A(r^2 \mathcal{G} \mathcal{H} B' (\mathcal{F} \mathcal{G} A'+\mathcal{J}^2 B')+B (\mathcal{F}
   \mathcal{G} (r^2 \mathcal{G} (\mathcal{H} A''+A' \mathcal{H}')-8 \mathcal{J}^2)\notag\\
   &-r^2 (\mathcal{G}^2 \mathcal{H} A' \mathcal{F}'+\mathcal{H}
   \mathcal{J}^2 B' \mathcal{G}'+\mathcal{G} \mathcal{J} (\mathcal{H} \mathcal{J} B''+\mathcal{J} B' \mathcal{H}'-2 \mathcal{H} B'
   \mathcal{J}')))\notag\\
   &+2 B^2 \mathcal{J} (\mathcal{G} (r (2 r \mathcal{H}' \mathcal{J}'-\mathcal{J} (r \mathcal{H}''+2
   \mathcal{H}'))+2 \mathcal{H} (2 r \mathcal{J}'+\mathcal{J}))-r \mathcal{J} \mathcal{G}' (r \mathcal{H}'+2 \mathcal{H})))).
\end{align}
\end{widetext}

\section{Analysis in the Eddington-Finkelstein coordinates}

To show that the instability we find in the main text is
not a coordinate artifact, we give a stability analysis
in a different coordinate system.
Here we work in the ingoing Eddington-Finkelstein coordinates $(v,r)$
that are used to see the regularity
of the horizons of the hairy black hole solutions~\cite{Babichev, Kobayashi:exa}.

The null coordinate $v$ is defined as
\begin{align}
v=t+\int^r\frac{d r'}{\sqrt{AB}}.
\end{align}
The metric perturbations are now given by
\begin{align}
h_{\mu\nu} &= h_{ta}dtdx^a+h_{ra}drdx^a
\notag \\
&=h_{ta} dvdx^a+\left(h_{ra}-\frac{h_{ta}}{\sqrt{AB}}\right)drdx^a,
\end{align}
in the $h_{ab}=0$ gauge. Let us denote the $(l,m)$ components of
$h_{va}:=h_{ta}$ and $h_{ra,{\rm EF}}:=h_{ra}-h_{ta}/\sqrt{AB}$ as
$\widetilde h_0(v,r)$ and $\widetilde h_1(v,r)$.
By an explicit calculation we obtain the quadratic action,
\begin{align}
S_{\rm EF}^{(2)}=\int dvdr{\cal L}_{\rm EF}^{(2)},
\end{align}
with
\begin{align}
\frac{2l+1}{2\pi}\mathcal{L}^{(2)}_{\rm EF}=&\; 
\widetilde a_1 \widetilde h_0^2+\widetilde a_2\widetilde h_1^2+\widetilde a_3
\biggl(\dot{\widetilde{h}}_1^2-2
   \widetilde h_0' \dot{\widetilde{h}}_1\notag\\
   & +\widetilde h_0'^2+\frac{4
   \widetilde h_0
   \dot{\widetilde{h}}_1}{r}\biggr)+\widetilde a_4
   \widetilde h_0\widetilde h_1.\label{eq:quadlagef}
\end{align}
Here a dot stands for $\partial_v$.
The coefficients are given by
\begin{align}
\widetilde a_1:=a_1+\frac{a_2}{AB}+\frac{a_4}{\sqrt{AB}},
\quad
\widetilde a_4:=a_4+\frac{2a_2}{\sqrt{AB}},\label{newcoeff}
\end{align}
while $a_2$ and $a_3$ remain the same, $\widetilde a_2=a_2$, $\widetilde a_3=a_3$.
The Lagrangian can also be derived just by
substituting $h_0\to \widetilde h_0$, $h_1\to\widetilde h_1+\widetilde h_0/\sqrt{AB}$,
$\partial_t\to \partial_v$, and $\partial_r\to\partial_r+(1/\sqrt{AB})\partial_v$
in Eq.~(\ref{eq:quadlag}). Thus, the quadratic Lagrangian in the Eddington-Finkelstein coordinates
has the same form as that in the $(t,r)$ coordinates derived in the main text,
but with the different coefficients.
The shift of the coefficients presented in (\ref{newcoeff}) is equivalent to redefining
\begin{align}
{\cal F}&\to\widetilde{\cal F}={\cal F}-{\cal G}+2\sqrt{\frac{B}{A}}{\cal J},
\\
{\cal J}&\to\widetilde{\cal J}={\cal J}-\sqrt{\frac{A}{B}}{\cal G},
\end{align}
while retaining the same ${\cal G}$ and ${\cal H}$.

Now, in the vicinity of the horizon, $A\ll 1$, we have
\begin{align}
\widetilde {\cal F}\simeq -\frac{2A}{q^2}X^2G_{4X},
\quad
{\cal G}\simeq +\frac{2q^2}{A}G_{4X},
\end{align}
showing that the black hole is unstable.



\begin{thebibliography}{99}
%
\bibitem{Israel:1967wq} 
  W.~Israel,
  ``Event horizons in static vacuum space-times,''
  Phys.\ Rev.\  {\bf 164}, 1776 (1967).

\bibitem{Israel:1967za} 
  W.~Israel,
  ``Event horizons in static electrovac space-times,''
  Commun.\ Math.\ Phys.\  {\bf 8}, 245 (1968).



\bibitem{Bekenstein1}
J.~D.~Bekenstein,
	``Black Holes: Classical Properties, Thermodynamics and Heuristic Quantization,''
[arXiv:gr-qc/9808028].

\bibitem{Bardeen:1973gs} 
  J.~M.~Bardeen, B.~Carter and S.~W.~Hawking,
  ``The Four laws of black hole mechanics,''
  Commun.\ Math.\ Phys.\  {\bf 31}, 161 (1973).

\bibitem{Kormendy:1995er} 
  See, {\em e.g.}, J.~Kormendy and D.~Richstone,
  ``Inward bound: The Search for supermassive black holes in galactic nuclei,''
  Ann.\ Rev.\ Astron.\ Astrophys.\  {\bf 33}, 581 (1995).
  


\bibitem{expan1}
S.~Perlmutter {\it et al.} [Supernova Cosmology Project Collaboration],
  ``Measurements of Omega and Lambda from 42 high redshift supernovae,''
  Astrophys.\ J.\  {\bf 517}, 565 (1999),
  [arXiv:astro-ph/9812133].
 \bibitem{Riess:1998cb} 
  A.~G.~Riess {\it et al.} [Supernova Search Team Collaboration],
  ``Observational evidence from supernovae for an accelerating universe and a cosmological constant,''
  Astron.\ J.\  {\bf 116}, 1009 (1998),
  [astro-ph/9805201].
  
\bibitem{Copeland:2006wr} 
  See, {\em e.g.}, E.~J.~Copeland, M.~Sami and S.~Tsujikawa,
  ``Dynamics of dark energy,''
  Int.\ J.\ Mod.\ Phys.\ D {\bf 15}, 1753 (2006),
  [hep-th/0603057].

\bibitem{clifton}
See, {\em e.g.}, T.~Clifton,~P.~G.~Ferreira,~A.~Padilla and C.~Skordis,
``Modified gravity and cosmology,''
Phys.\ Rep.,\ {\bf 513},\ 1  (2012), [arXiv:1106.2476 [astro-ph.CO]].

\bibitem{Hawking:1972qk} 
  S.~W.~Hawking,
  ``Black holes in the Brans-Dicke theory of gravitation,''
  Commun.\ Math.\ Phys.\  {\bf 25}, 167 (1972).


\bibitem{Sotiriou:2011dz} 
  T.~P.~Sotiriou and V.~Faraoni,
  ``Black holes in scalar-tensor gravity,''
  Phys.\ Rev.\ Lett.\  {\bf 108}, 081103 (2012),
  [arXiv:1109.6324 [gr-qc]].
  
\bibitem{Graham:2014mda} 
  A.~A.~H.~Graham and R.~Jha,
  ``Nonexistence of black holes with noncanonical scalar fields,''
  Phys.\ Rev.\ D {\bf 89}, 084056 (2014),
  [Phys.\ Rev.\ D {\bf 92},  069901 (2015)],
  [arXiv:1401.8203 [gr-qc]].
  
\bibitem{Graham:2014ina} 
  A.~A.~H.~Graham and R.~Jha,
  ``Stationary Black Holes with Time-Dependent Scalar Fields,''
  Phys.\ Rev.\ D {\bf 90}, no. 4, 041501 (2014)
  [arXiv:1407.6573 [gr-qc]].

\bibitem{Hui}
L.~Hui~and~A.~Nicolis,
``A no-hair theorem for the galileon,"
Phys.\ Rev.\ Lett.\ {\bf 110},\ 241104 (2013), 
[arXiv:1202.1296 [hep-th]].

\bibitem{Herdeiro:2015waa} 
  C.~A.~R.~Herdeiro and E.~Radu,
  ``Asymptotically flat black holes with scalar hair: a review,''
  Int.\ J.\ Mod.\ Phys.\ D {\bf 24}, no. 09, 1542014 (2015),
  [arXiv:1504.08209 [gr-qc]].

\bibitem{Babichev}
E.~Babichev~and~C.~Charmousis,
	``Dressing a black hole with a time-dependent Galileon,''
JHEP {\bf 1408} (2014) 106,
[arXiv:1312.3204 [gr-qc]].


\bibitem{Rinaldi}	
M.~Rinaldi,
 ``Black holes with non-minimal derivative coupling,''
  Phys.\ Rev.\ D {\bf 86}, 084048 (2012).

\bibitem{Anabalon:2013oea} 
  A.~Anabalon, A.~Cisterna and J.~Oliva,
  ``Asymptotically locally AdS and flat black holes in Horndeski theory,''
  Phys.\ Rev.\ D {\bf 89}, 084050 (2014),
  [arXiv:1312.3597 [gr-qc]].
 
\bibitem{Minamitsuji1}
M.~Minamitsuji,
	``Solutions in the scalar-tensor theory with nonminimal derivative coupling,''
Phys.\ Rev.\ {\bf D89},\ 064017 (2014), [arXiv:1312.3759 [gr-qc]].

\bibitem{Sotiriou1}
T.~P.~Sotiriou and S.-Y.~Zhou,
	``Black hole hair in generalized scalar-tensor gravity,''
Phys.\ Rev.\ Lett.\ {\bf 112},\ 251102 (2014),
[arXiv:1312.3622 [gr-qc]].

\bibitem{Sotiriou2}
T. P. Sotiriou and S.-Y. Zhou, 
	``Black hole hair in generalized scalar-tensor gravity: An explicit example,''
Phys.\ Rev.\ {\bf D90},\ 124063 (2014),
[arXiv:1408.1698 [gr-qc]].



\bibitem{Kobayashi:exa}
T.~Kobayashi~and~N.~Tanahashi,
	``Exact black hole solutions in shift-symmetric scalar-tensor theories,''
PTEP~{\bf 2014},~073E02 (2014),
[arXiv:1403.4364 [gr-qc]].

\bibitem{Horndeski}
  G.~W.~Horndeski,
  ``Second-order scalar-tensor field equations in a four-dimensional space,''
  Int.\ J.\ Theor.\ Phys.\  {\bf 10}, 363 (1974).
  
\bibitem{Deffayet} 
  C.~Deffayet, X.~Gao, D.~A.~Steer and G.~Zahariade,
  ``From k-essence to generalised Galileons,''
  Phys.\ Rev.\  {\bf D84}, 064039 (2011),
  [arXiv:1103.3260 [hep-th]].

\bibitem{Kobayashi:ge} 
  T.~Kobayashi, M.~Yamaguchi and J.~Yokoyama,
  ``Generalized G-inflation: Inflation with the most general second-order field equations,''
  Prog.\ Theor.\ Phys.\  {\bf 126}, 511 (2011),
  [arXiv:1105.5723 [hep-th]].

\bibitem{Charmousis:2014zaa} 
  C.~Charmousis, T.~Kolyvaris, E.~Papantonopoulos and M.~Tsoukalas,
  ``Black Holes in Bi-scalar Extensions of Horndeski Theories,''
  JHEP {\bf 1407}, 085 (2014),
  [arXiv:1404.1024 [gr-qc]].

\bibitem{Babichev:2015rva} 
  E.~Babichev, C.~Charmousis and M.~Hassaine,
  ``Charged Galileon black holes,''
  JCAP {\bf 1505}, 031 (2015)
  [arXiv:1503.02545 [gr-qc]].

\bibitem{DeFelice}
A.~De~Felice,~T.~Suyama,~and~T.~Tanaka,
	``Stability of Schwarzschild-like solutions in f(R,G) gravity models,''
Phys.\ Rev.\ {\bf D83},\ 104035 (2011),
[arXiv:1102.1521 [gr-qc]].

\bibitem{Motohashi1}
H.~Motohashi~and~T.~Suyama,
	``Black hole perturbation in non-dynamical and dynamical Chern-Simons gravity,''
Phys.\ Rev.\ {\bf D85} (2012) 044054, 
[arXiv:1110.6241 [gr-qc]].

\bibitem{Motohashi:2011pw} 
  H.~Motohashi and T.~Suyama,
  ``Black hole perturbation in parity violating gravitational theories,''
  Phys.\ Rev.\ D {\bf 84}, 084041 (2011),
  [arXiv:1107.3705 [gr-qc]].


\bibitem{Kobayashi:per1}
T.~Kobayashi,~H.~Motohashi,~and~T.~Suyama,
	``Black hole perturbation in the most general scalar-tensor theory with second-order field equations I: the odd-parity sector,''
Phys.\ Rev.\ {\bf D85},\ 084025 (2012),
[arXiv:1202.4893 [gr-qc]].

\bibitem{Kobayashi:per2}
T.~Kobayashi,~H.~Motohashi,~and~T.~Suyama, 
	``Black hole perturbation in the most general scalar-tensor theory with second-order field equations II: the even-parity sector,''
Phys.\ Rev.\ {\bf D89},\ 084042 (2014),
[arXiv:1402.6740 [gr-qc]].

\bibitem{Cisterna:2015uya} 
  A.~Cisterna, M.~Cruz, T.~Delsate and J.~Saavedra,
  ``Nonminimal derivative coupling scalar-tensor theories: odd-parity perturbations and black hole stability,''
[arXiv:1508.06413 [gr-qc]].


\bibitem{Germani}
C.~Germani and A.~Kehagias,
``Cosmological Perturbations in the New Higgs Inflation,''
JCAP 1005, 019 (2010) [Erratum-ibid. 1006, E01 (2010)], [arXiv:1003.4285 [astro-ph.CO]].

  \bibitem{Gubitosi:2011sg} 
  G.~Gubitosi and E.~V.~Linder,
  ``Purely Kinetic Coupled Gravity,''
  Phys.\ Lett.\ B {\bf 703}, 113 (2011),
  [arXiv:1106.2815 [astro-ph.CO]].


\bibitem{Regge}
T.~Regge~and~J.~A.~Wheeler,
	``Stability of a Schwarzschild Singularity,''
Phys.\ Rev.\ {\bf 108},\ 1063 (1957).

\bibitem{Zerilli}
F.~J.~Zerilli, 
	``Effective Potential for Even-Parity Regge-Wheeler Gravitational Perturbation Equations,''
Phys.\ Rev.\ Lett. {\bf 24} , 737 (1970).

\bibitem{Moncrief:1974am} 
  V.~Moncrief,
  ``Gravitational perturbations of spherically symmetric systems. I. The exterior problem.,''
  Annals Phys.\  {\bf 88}, 323 (1974).
  
  
\bibitem{Maselli:2015yva} 
A.~Maselli, H.~O.~Silva, M.~Minamitsuji and E.~Berti,
  ``Slowly rotating black hole solutions in Horndeski gravity,''
  Phys.\ Rev.\ D {\bf 92}, 104049 (2015)
  [arXiv:1508.03044 [gr-qc]].

\bibitem{Woodard:2015zca} 
R.~P.~Woodard,
  ``Ostrogradsky's theorem on Hamiltonian instability,''
  Scholarpedia {\bf 10}, 32243 (2015)
  [arXiv:1506.02210 [hep-th]].
\end{thebibliography}
\end{document}